\documentclass[nonlinenumbers]{aastex631}

\shorttitle{The PN in Open Cluster M37}
\shortauthors{Fragkou et al.}
\graphicspath{{./}{figures/}}
\begin{document}

\title{The Planetary Nebula in the 500~Myr old Open Cluster M37}

\correspondingauthor{Quentin A. Parker}
\email{quentinp@hku.hk}

\author[0000-0003-0869-4847]{Vasiliki Fragkou}
\affiliation{Instituto de Astronomía, Universidad Nacional Autónoma de México, 22800, Ensenada, B.C., Mexico}

\author[0000-0002-2062-0173]{Quentin A. Parker}
\affiliation{Department of Physics, The University of Hong Kong, Chong Yuet Ming Physics Building, Pokfulam Road, Hong Kong}
\affiliation{Laboratory for Space Research, The University of Hong Kong, 405B Cyberport 4, 100 Cyberport Road, Cyberport, Hong Kong}

\author[0000-0002-3171-5469]{Albert A. Zijlstra}
\affiliation{Jodrell Bank Centre for Astrophysics, The University of Manchester, Oxford Road, M13 9PL, Manchester, UK}
\affiliation{Laboratory for Space Research, The University of Hong Kong, 405B Cyberport 4, 100 Cyberport Road, Cyberport, Hong Kong}

\author[0000-0002-3279-9764]{Roberto Vázquez}
\affiliation{Instituto de Astronomía, Universidad Nacional Autónoma de México, 22800, Ensenada, B.C., Mexico}

\author[0000-0003-0242-0044]{Laurence Sabin}
\affiliation{Instituto de Astronomía, Universidad Nacional Autónoma de México, 22800, Ensenada, B.C., Mexico}

\author[0000-0002-0121-2537]{Jackeline Suzett Rechy-Garcia }
\affiliation{Instituto de Radioastronomía y Astrofísica, Universidad Nacional Autónoma de México, 58089, Morelia, Michoacan, Mexico}

%% Note that the \and command from previous versions of AASTeX is now
%% depreciated in this version as it is no longer necessary. AASTeX 
%% automatically takes care of all commas and "and"s between authors names.

%% AASTeX 6.31 has the new \collaboration and \nocollaboration commands to
%% provide the collaboration status of a group of authors. These commands 
%% can be used either before or after the list of corresponding authors. The
%% argument for \collaboration is the collaboration identifier. Authors are
%% encouraged to surround collaboration identifiers with ()s. The 
%% \nocollaboration command takes no argument and exists to indicate that
%% the nearby authors are not part of surrounding collaborations.

%% Mark off the abstract in the ``abstract'' environment. 
\begin{abstract}

We report confirmation of a large, evolved, 
bipolar planetary nebula and its blue, white dwarf 
central star as a member of the $\sim$500~Myr old 
Galactic open star cluster M37 (NGC~2099). This is only the third known 
example of a planetary nebula in a Galactic open cluster and was found via our 
on-going program of identifying and studying planetary nebulae - 
open cluster associations. High confidence in 
the association  comes from the
consistent radial velocities and proper motions
for the confirmed central star and cluster stars from Gaia,
reddening agreement and
location of the planetary nebula well within the 
tidal cluster boundary. 
Interestingly, all three Galactic examples have bipolar morphology
and likely Type~I chemistry, both characteristics 
of higher mass progenitors. In this case the 
progenitor star mass is in the mid-range of 
$\sim$2.8~M$_\sun$. It provides a valuable, additional 
point on the key stellar initial-to-final mass 
relation independent
of cluster white dwarf estimates and also falls in a gap in the poorly sampled mass region. This planetary nebula also appears to have the largest kinematical age ever determined and implies increased visibility lifetimes when they are located in clusters.
\end{abstract}

%% Keywords should appear after the \end{abstract} command. 
%% The AAS Journals now uses Unified Astronomy Thesaurus concepts:
%% https://astrothesaurus.org
%% You will be asked to selected these concepts during the submission process
%% but this old "keyword" functionality is maintained in case authors want
%% to include these concepts in their preprints.

%% From the front matter, we move on to the body of the paper.
%% Sections are demarcated by \section and \subsection, respectively.
%% Observe the use of the LaTeX \label
%% command after the \subsection to give a symbolic KEY to the
%% subsection for cross-referencing in a \ref command.
%% You can use LaTeX's \ref and \label commands to keep track of
%% cross-references to sections, equations, tables, and figures.
%% That way, if you change the order of any elements, LaTeX will
%% automatically renumber them.
%%
%% We recommend that authors also use the natbib \citep
%% and \citet commands to identify citations.  The citations are
%% tied to the reference list via symbolic KEYs. The KEY corresponds
%% to the KEY in the \bibitem in the reference list below. 

\section{Introduction} \label{sec:intro}

Planetary nebulae (PNe) are a brief, typically $\sim$5,000--25,000 year 
\citep{2015ApJ...804L..25B}, phase of stellar evolution. The central star (CSPN) is the 
low-mass core of the previous  Asymptotic Giant Branch (AGB) star, which has expelled its 
envelope. The CSPN quickly
evolves to become a white dwarf (WD) on the cooling track. If the stellar age is known (i.e. 
the AGB turn-off age for its initial mass), the object can constrain the initial-final mass 
relation (IFMR).

Planetary nebulae provide us with vital clues for 
understanding late stage stellar evolution and Galactic chemical 
enrichment. Their strong emission lines allow 
determination of abundances, expansion and radial velocities and 
CSPN temperatures. PNe yield information on 
the nuclear burning, dredge up and mass loss in the stellar 
progenitor. See \citet{2022PASP..134b2001K} for an 
excellent recent PN review.  PN studies have been hampered by three 
problems: i) the previous lack of accurate distances to most 
Galactic PNe; ii) obtaining representative PNe samples 
of the true population diversity
\citep{2022FrASS...9.5287P} and iii) their unknown progenitor 
masses. The first problem has prospects of resolution via 
accurate Gaia\footnote{https://www.cosmos.esa.int/web/gaia/dr3} CSPN distances, 
though many 
CSPN remain too distant and faint for Gaia DR3 and correct CSPN identification 
remains an issue for some \citep{2022Galax..10...32P}. The second problem is being 
addressed by deep, narrow-band, wide-field surveys, e.g. 
\citet{2005MNRAS.362..689P}, \citet{2005MNRAS.362..753D} and 
\citet{2014MNRAS.440.2036D}. 

For the third problem of progenitor masses, these 
can only be accurately determined for PNe in 
Galactic Globular and open clusters (OCs). These allow precise distance 
determinations from color-magnitude diagrams (CMD) and Gaia. 
Importantly, they provide good progenitor mass estimates from the 
cluster isochrones and the measured cluster turn-off mass. A proven physical association of 
an OC with a PN is an extremely valuable probe. It can 
provide us with: i) accurate physical characteristics of the 
PN and its CSPN from the known distance; ii) the metallicity from the nebula and cluster and iii) the age and mass of the 
progenitor star. We can then study the relation between stellar mass 
and the PN's chemical enrichment, determined from 
spectroscopic measurements, and provide additional, independent 
data for the widely used WD IFMR,
e.g. \citet{2009MNRAS.395.2248D}, that 
associates WD properties to their main-sequence progenitors. A well
determined IFMR is crucial for tracing the development of both 
carbon and nitrogen in entire galaxies but remains currently 
poorly constrained.

There are currently only four confirmed PNe in Galactic globular 
clusters \citep{1997AJ....114.2611J} but see \citet{2020AJ....159..276B} where Gaia proper motions bring doubt to PN JaFu1
being a member of globular cluster Palomar~6.  
\citet{2019ApJ...884L..15M} proposed four 
additional PNe candidates identified in several of the $\sim$50 new globular
clusters found in the Galactic bulge from the VVV survey \citep{2012A&A...537A.107S}
but substantial follow-up is needed to confirm any of these. 

For OCs on the other 
hand, before this work, only 2 PNe had been confirmed as physically associated with Galactic
OCs (PN PHR 1315$-$6555 in the OC Andrews-Lindsay\,1, and PN BMP J1613$-$5406 in the OC NGC\,6067), both discovered by members of  our team: \cite{2011MNRAS.413.1835P}  and \citet{fragkou2019}.\footnote{BMP J1613$-$5406's association with NGC\,6067 is based on radial velocity, reddening and distance agreement, location of the PN within the cluster's tidal radius and other considerations. The GAIA DR3 proper motion is inconclusive on this source, with the proper motion differing from that of the cluster by 2.3$\sigma$, however the slightly higher RUWE (renormalized error) of 1.3 indicates an enhanced uncertainty and the star is very close to the Gaia magnitude limits. }

\citet{2019ApJ...884..115D} detected a PN that may be a physical member of an OC in 
the M31 local group galaxy indicating the first discovery of an extragalactic PN-OC 
pair. They estimate a PN initial mass of 3.38$_{-0.02}^{+0.03}$ M$_\sun$ and enhanced nitrogen
abundance, indicating that hot bottom burning can be evident in relatively low mass AGB stars.
The system's remoteness makes detailed study and robust confirmation requirements, as applied 
to Galactic counterparts, challenging.

As part of a recent astro-photometric 
study of M37, \citet{https://doi.org/10.48550/arxiv.2207.03179}
found that a WD, identified as a high-probability proper motion member of the cluster, is also the likely 
central star of a previously known PN detected in the IPHAS H$\alpha$ survey \citep{2005MNRAS.362..753D}.  This association was first proposed by 
\citet{2017IAUS..323...11F} and subsequently studied by the first author
\citet{2019PhDT........68F} in the context of a 
possible association with M37. This was based on nebular emission 
contamination in the fibre spectra of several M37
cluster stars \citet{2017ApJ...834..176N} and the 
presence of an extended, diffuse emission region near the cluster
core in a mosaic of the relevant IPHAS H$\alpha$ images
(see their Fig.\,4). We prove here the association of the PN with M37 and so add an 
important object to a very small sample.

\section{Association of PN IPHASX~J055226.2+323724 with Galactic Open Cluster M37}

\subsection{The PN}

The nebula was discovered and classified as a PN candidate by \citet{Sabin2008} from IPHAS imagery.  Here we present new, high resolution radial velocity data 
that confirm this rare OC-PN link. The PN is identified as a `True', bipolar, likely Type-I
PN in HASH \citep{2016JPhCS.728c2008P} as IPHASX~J055226.2+323724 
(PNG~177.5+03; HASH ID~31188) where a clearer image than that
presented in \citet{2017ApJ...834..176N} shows its 
evolved, bipolar nature. It is of very low surface brightness at the 
$\sim$5 Rayleigh sensitivity limit of IPHAS 
\citep{2005MNRAS.362..753D} with a major axis of 
445$\pm$10~arseconds. The PN presents enhancements along the southern edges and a patchy
internal structure (see fig.\,1).  The emission line 
spectra seen in the cluster stellar fibre-spectra for five 
stars in Fig.\,5 of \citep{2017ApJ...834..176N} show high [\ion{N}{2}]/H$\alpha$ ratios, indicative of a  likely Type-I (nitrogen enriched) chemistry 
\citep{1994MNRAS.271..257K}.

\begin{figure*}[tbph]
	\centering
\includegraphics[width=12cm]{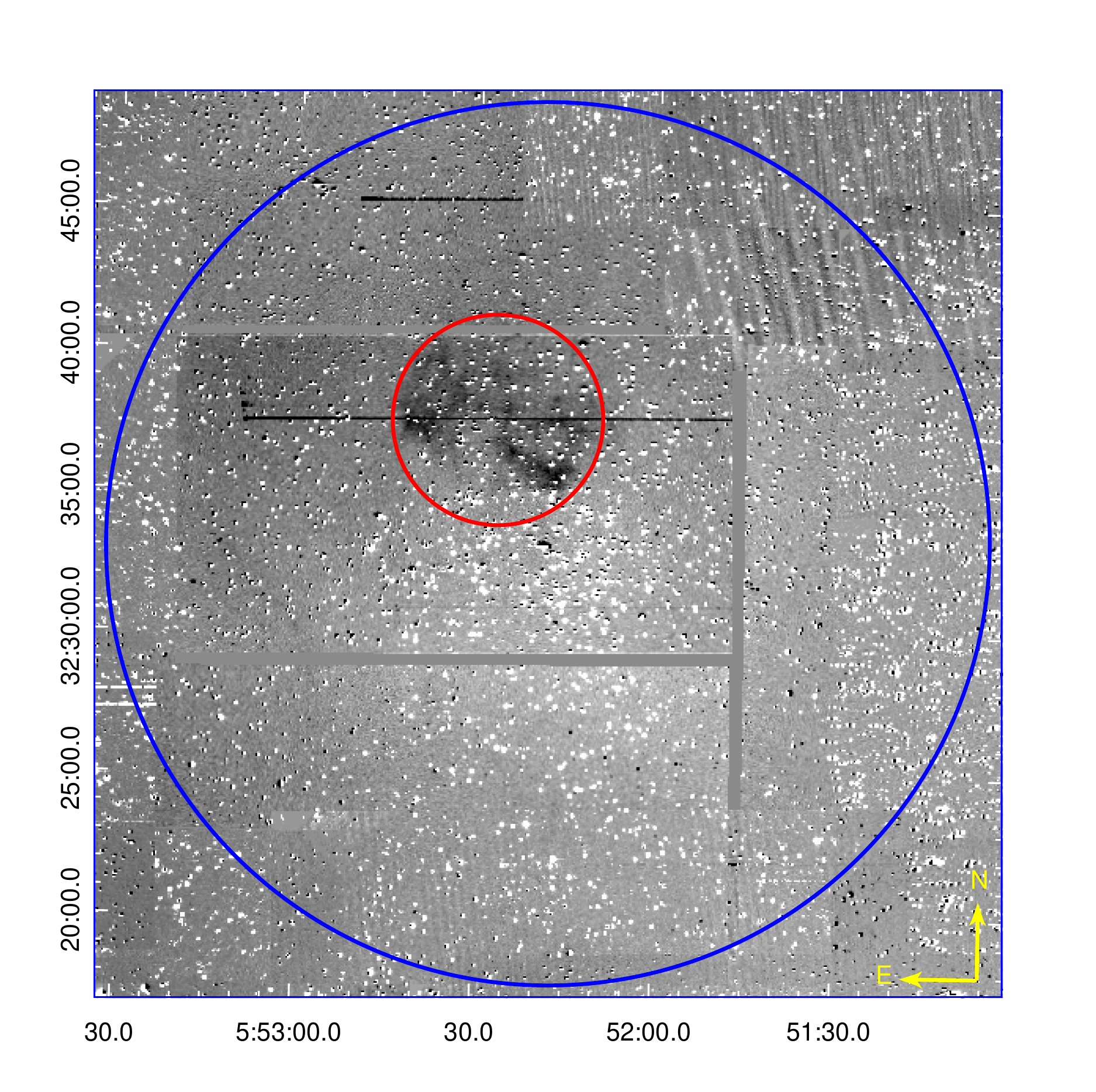}
	\caption{A contrast enhanced 30$\times$30~arcminute quotient (H$\alpha$\,$-$\,r band) IPHAS \citep{2005MNRAS.362..753D} mosaic centred on the core of Galactic open cluster M~37 (NGC~2099). The low surface brightness bipolar PN (IPHASX~J055226.2+323724) is encompassed by a red circle with a diameter of 445~$\pm$~10~arcseconds (the nebular major axis) while the blue circle indicates the full $\sim30$~arcminute extent of the cluster. The PN is well within the cluster tidal radius with the blue CSPN at almost the precise geometric centre of the PN. The CSPN is itself only $\sim$280~arcseconds from the published cluster center position.
	}
	\label{fig:colour}
\end{figure*}

\subsection{The CSPN}
Based on Sloan Digital Sky Survey imagery 
\citep{2006AJ....131.2332G} we found the CSPN at RA: 
05:52:26.18, DEC: 32:37:24.63 (J2000), almost exactly at the PN's
geometric centre ($<$10~arcsecond 
displacement) which is $\sim$7.2~arcmin across its major axis. This is also the 
only blue star within a 116~arcsecond radius from the PN's center, making it the only 
plausible CSPN candidate. This CSPN was previously 
identified as a WD candidate in \citet{2015MNRAS.448.2260G}. It is also reported as
a CSPN by \citet{2020A&A...638A.103C} via Gaia 
searches for CSPN based on PN in the HASH catalog \citep{2016JPhCS.728c2008P}, 
though no association with the cluster was made. 

Low dispersion spectroscopy of the CSPN is presented in 
\citet{https://doi.org/10.48550/arxiv.2207.03179} that show both He~II  and C~IV absorption lines, confirming the CSPN is a 
hot, hydrogen deficient WD. It is reported as an intermediate type between the DO class and 
the PG1159 stars, e.g. 
\citet{2006PASP..118..183W}
and \citet{2014A&A...572A.117R}. They report a stellar
$T_{eff}>60,000$K. (We estimate a value closer to 100,000~K in this paper). Hence, this CSPN 
is hot enough to ionise the observed PN. The Gaia EDR3 ID of
this CSPN is 3451205783698632704.

\begin{figure}[h!]
\begin{center}
\includegraphics[width=16cm]{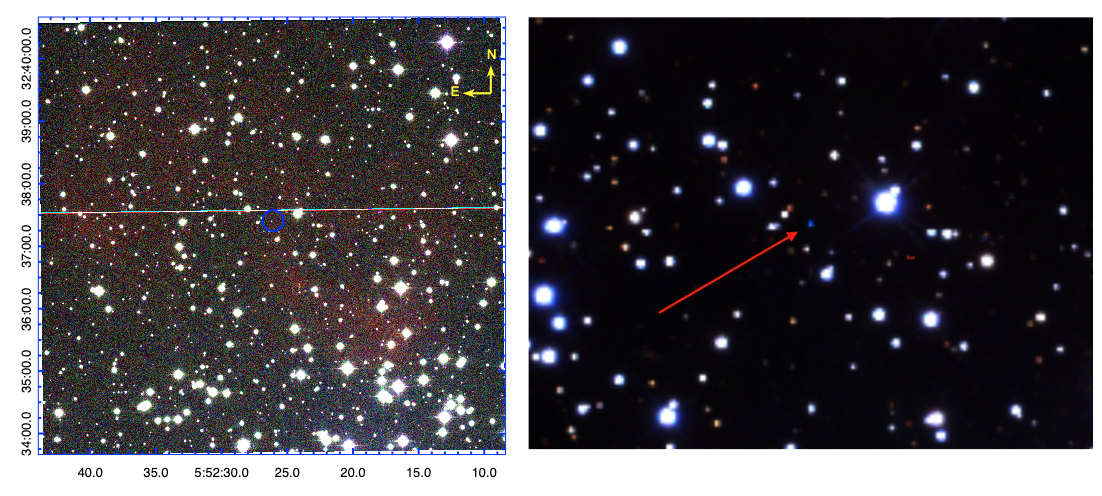}
\end{center}
\caption{Fig.\,2a left panel: An enhanced 6.5$\times$6.5~arcminute colour-composite RGB image
of PN IPHASX J055226.2+323724 from the IPHAS survey 
\citep{2005MNRAS.362..753D} that we confirm as a physical member 
of the Galactic open cluster M37. Red = H$\alpha$, Green = broad band red and Blue = broad band `i'. The CSPN is circled in blue; Fig.\,2b right panel: 
190$\times$145 arcsecond RGB image created from SDSS with red = i, green = r and 
blue = g-band. These data clearly shows the faint CSPN (arrowed) at  the
centre. North is top and East is to the left in both 
images.}\label{fig:CDM-IFMR}
\end{figure}

The white dwarf has been classified as a rotating variable with a period of 
0.445 days and amplitude of 0.074 mags by \citet{Chang2015} (V1975 in their sample). Rotational variability has previously been found in hot DQ white dwarfs \citep{Lawrie2013, Williams2016} where it is related to strong magnetic fields. The current star is not known to be of DQ type but this raises the possibility that it could be a transition object between PG1159 stars and DQ white dwarfs. Alternatively, the variability could be due to a close binary, however it is not classified as such by \citet{Chang2015}.

%\newpage
\subsection{The cluster M37 (NGC~2099)}
M37 (NGC~2099) is the brightest and richest Galactic open cluster in the  
constellation of Auriga with a stellar mass  of $\sim$1500\,M$_\odot$  based on concordant 
Gaia EDR3 \citep{2021A&A...649A...1G} high probability proper motions and potentially with as
many as 4500 stars. It is a well-studied, intermediate-age ($\sim$500 million year old)
cluster first discovered by the Italian astronomer Giovanni Battista 
Hodierna before 1654 and catalogued by Messier in 1781. It is about 
30~arcminutes across at full extent. Previous modern work includes 
\citet{2001AJ....122.3239K}, 
\citet{2005ApJ...618L.123K} and 
\citet{2007A&A...463..789A} who show
the cluster contains three blue stragglers while its hottest surviving 
main sequence star is a B9V. More recent work on M~37 concentrates on the 
Gaia data, e.g. \citet{2022MNRAS.511.4702G} 
and \citet{https://doi.org/10.48550/arxiv.2207.03179} who also list 7 
cluster WD candidates including the CSPN labelled WD\,1 in their list. The weighted average cluster 
physical parameters obtained from the literature have been previously estimated as: angular diameter
= 31.15 arcminutes, age = 470~$\pm$~50 Myrs, reddening $E(B-V)$ = 0.26$~\pm$~0.04, distance = 
1.44~$\pm$~0.13 kpc and metallicity $[Fe/H]$ = 0.03~$\pm$~0.28 
(\citet{2022MNRAS.511.4702G}; refer table 6.6 in \citet{2019PhDT........68F}). Using 1136 stars with a cluster membership probability $>$
0.8, \citet{2018A&A...618A..93C} yields a Gaia DR3 mean 
distance of 1485~pc with $\sigma$ = 
110~pc, in good agreement with previous estimates. Since this is based on 
the latest Gaia DR3 data, we are using this cluster distance value throughout. The same stars 
present a 
Gaia mean pmRA$\, = 1.88\pm 0.18$ mas/yr and pmDec$\, = -5.62 \pm 0.16$ mas/yr. The Gaia DR3
CSPN proper motion is in tight agreement with the cluster average, strongly supporting 
cluster membership (Table. 1). The parallax has a larger uncertainty but is 
consistent within $1\sigma$. For the adopted cluster distance and angular diameter, the 
cluster's physical radius is 6.73~pc. 

\section{Methods}

\subsection{Magnitudes of the CSPN}
The ugriz CSPN magnitudes have been measured from SDSS DR10 
\citep{2007ApJ...671.1640A} as \emph{u} = 18.702 $\pm$ 0.020 mag, \emph{g} =
18.978 $\pm$ 0.010 
mag, \emph{r} = 19.300 $\pm$ 0.011 mag, \emph{i} = 20.248 $\pm$ 0.063 mag and \emph{z}= 
19.840~$\pm$~0.082 mag \citep{2015MNRAS.448.2260G}. Following 
\citet{2005AJ....130..873J} these are 
transformed to the Johnson-Cousins system as \emph{B} = 19.07 $\pm$ 0.02 and
\emph{V} = 19.16 $\pm$ 0.02. The errors have been computed with standard error propagation 
with the transformation rms residuals added quadratically. The CSPN is also included in the Pan-STARRS catalogue \citep{2016arXiv161205560C} with magnitudes $g=19.06\pm0.015$, $r= 19.40\pm0.01$,  $i= 19.69\pm0.02$, $z= 19.93\pm0.03$ and $v= 20.03\pm0.04$.

\subsection{Spectroscopic Observations of the PN}

For PN radial velocity measurement we obtained 
observations with the high spectral resolution MEGARA spectrograph 
\citep{2020MNRAS.493..871G} on 12$^{th}$ January 2022 (with full moon 
contamination) and 4$^{th}$ March 2022 (dark sky). MEGARA is an 
optical (3650–-9750\AA) medium-high spectral resolution fibre-fed 
spectroscopic instrument on the 10.4-m Gran Telescopio CANARIAS (GTC) in 
La Palma, Canary Islands. We obtained optical spectroscopy at 7 
distinct positions  across the PN (refer Fig.3a). Multiple 5-minute 
exposures were taken in IFU mode. The IFU is a bundle of 
567 (0.62 arcsec diameter) fibres
that subtend 12.5~$\times$~11.3~arcseconds on the sky. One pointing
was centered on the CSPN (pointing a), plus an offset sky exposure. The VPH
grating set up was the `HR--R' highest dispersion mode (R = 20,000) that 
covers 6400--6800\AA\, and gives a radial velocity 
precision of around 1 km/s, ideal for our purpose. Four areas (a, b, c and
d) of the total of the seven pointings were observed in both January and 
March 2022. The earlier data was affected by the Moon causing a strong 
H$\alpha$ absorption feature but the [NII] and [SII] lines were 
unaffected. The pointing map and an example combined nebula spectrum from 
pointings a, b, c, d (for observations made on March 4$^{th}$ 2022) are shown in Fig.\,3a,b.

\begin{figure}[h!]
\begin{center}
\includegraphics[width=8cm]{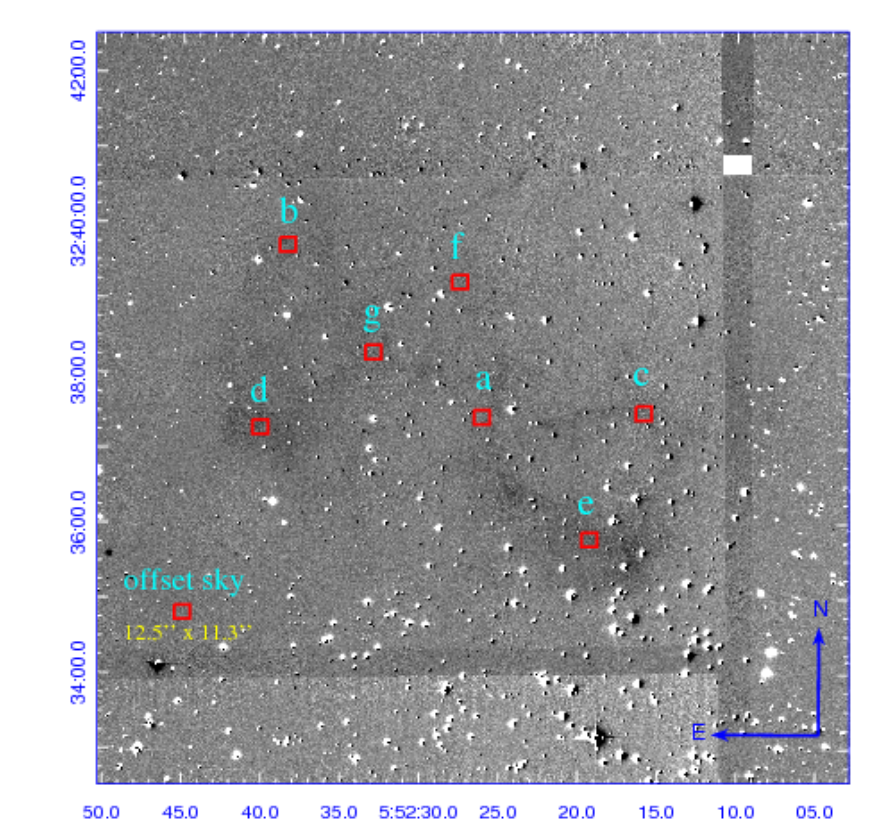}
\includegraphics[width=9.5cm, height=7.5cm]{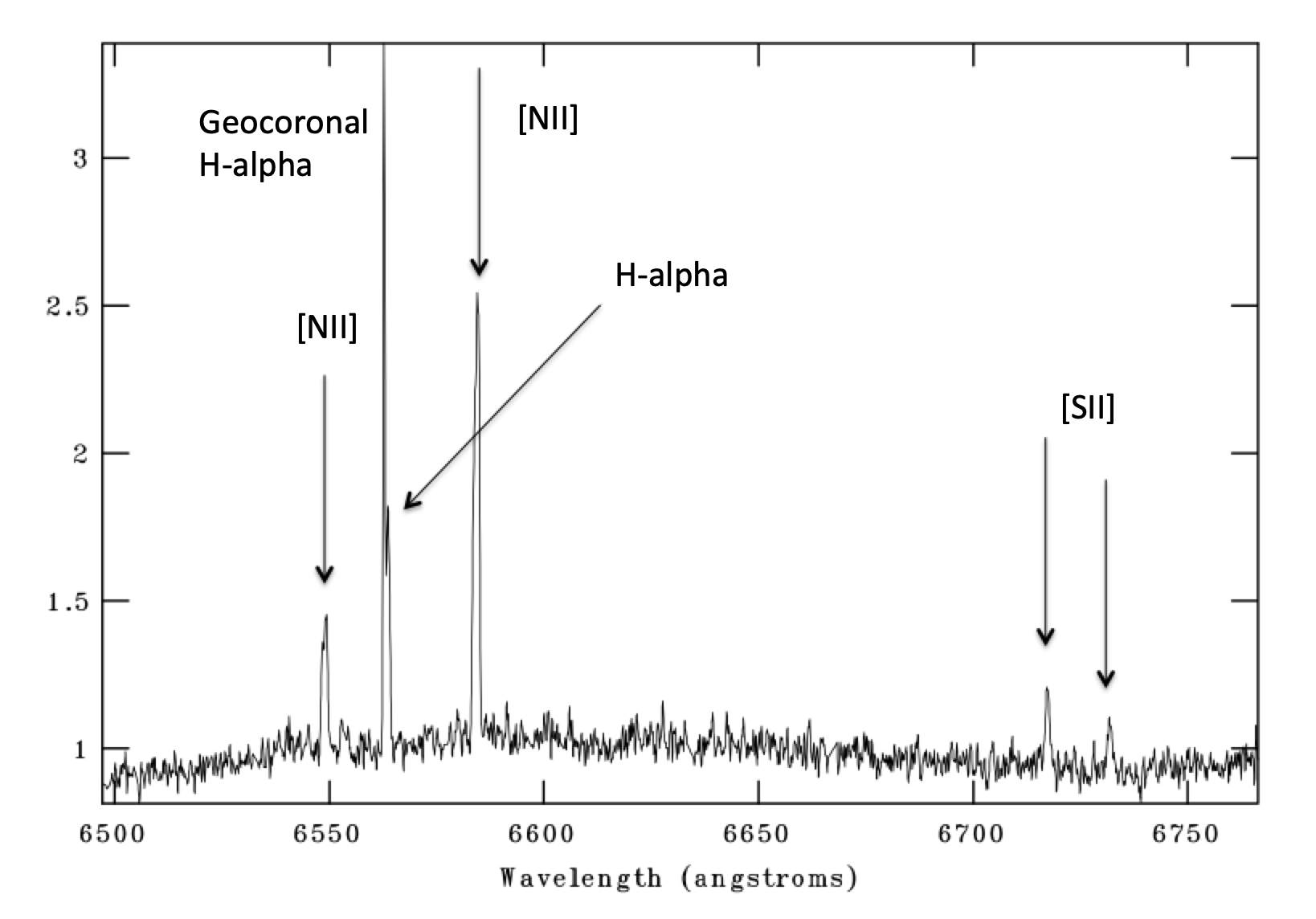}
\end{center}
\caption{Left panel: PN image with MEGARA IFU pointings 
indicated. Right panel: A combined 1-d 
continuum subtracted example PN spectrum from March 4$^{th}$ 2022 for IFU pointings a, b, c 
and d. The 5 visible PN emission lines are labelled.}\label{fig:CDM-IFMR}
\end{figure}

\subsection{Radial velocity, kinematic age and electron density estimates of the 
PN}\label{subsec:fitting}

The robust MEGARA pipeline was employed to reduce the data in standard fashion
for IFU mode. Flux calibration and accurate sky subtraction are not essential 
for the current purposes. The wavelength calibration appears robust. The radial 
velocities for each pointing are measured from the nebular emission lines in the integrated 
IFU spectrum of each pointing. After the standard reduction
with the MEGARA pipeline (which also combined the exposures for each observing
block), the integrated 1-D spectrum for each pointing was constructed using 
standard IRAF techniques. The S/N for one of our pointings (area f) from 
January 13$^{th}$ 2022 was too low for useful measurements. Otherwise the nebular emission 
lines were used for radial velocity determinations via Gaussian fitting. 
%and cross checked with cross correlation of all 1-D pointings. 

We computed the radial velocity average for all pointings with repeated 
observations (more than one observing block) and calculated the median heliocentric 
corrected radial velocity of all pointings as $v_{\rm rad} = 10.6 \pm  4.9$\,km/s from 6 
individual combined IFU pointings (4 were observed twice on different 
nights). The nebular spectra covered the [\ion{N}{2}], H$\alpha$ and [\ion{S}{2}] emission lines 
which were all detected. The weaker [\ion{N}{2}]\,6548\AA\, and [\ion{S}{2}]  doublet lines had too
low S/N in individual pointings to give reliable radial velocities due to larger errors.  The
6584\AA\, [\ion{N}{2}] line (from all individual observing blocks) and the H$\alpha$ line (from 
observations acquired in March and not affected by the strong H$\alpha$ absorption feature)
have both been employed for the calculation of the mean nebular radial 
velocity.  This value is compatible with the cluster heliocentric corrected velocity of 
$v_{\rm rad} = 8.32 \pm 0.56$\,km/s providing a tight constraint for cluster membership. 
In most pointings the [\ion{N}{2}] 6584\AA\, line was split into a blue-shifted and red-shifted 
component, with various asymmetries. allowing a direct measurement of the nebula expansion 
velocity at those pointings. The 
pointing `a' on the PN center provided the clearest splitting 
of the [\ion{N}{2}] 6584\AA\, line. These provided an average expansion velocity of 20 km/s  
with a standard deviation of 6.2 km/s. This is typical for a PN and allows a 
kinematic age to be determined from the PN physical size from its angular 
extent and distance, of  $t_{\rm kin}\sim 78\times 10^3 \pm25 \times 10^3$~yrs. This 
kinematic age is at the extreme end for PNe as befits such an evolved, large and low-surface 
brightness example. It may in fact be the largest PN kinematic age ever determined, assuming 
invariant expansion velocity over time. 

For the [\ion{S}{2}] line electron density estimate we combined the a,b,c,d
pointings from March 4$^{th}$ 2022 to provide a higher S/N,  1-D PN continuum 
subtracted spectrum. A box-3 smooth was applied before Gaussian fits to the
two well detected [\ion{S}{2}] lines using a wide wavelength range around the lines for determining 
the best base level. As expected the [\ion{S}{2}] lines are found to be in the 
low density limit with [\ion{S}{2}]6717/6731$\sim$1.45$~\pm~$0.20 from repeat measures. As such an 
electron density of $N_e < 5  cm^{-3}$, was obtained using the IRAF {\it nebular} add-on
package and assuming a $T_e = 10,000$\,K.  We use this value, the average of the nebular
minor and major physical radius, and an assumed filling
factor of 0.3 \citep{1996A&A...307..561P} to calculate a maximum ionised PN gas mass of 
0.32~M$_\sun$  \citep{1994A&A...284..248B}. From the same data we measured a 
[\ion{N}{2}]/H$\alpha$=$3.58 \pm 0.10$ which supports the Type~I nature of the bipolar 
nebula.

\subsection{Interstellar extinction}

O5 Main Sequence stars present a $B - V$ color of $- 0.33$ 
\citep{2000asqu.book.....C} and since 
hotter stars are expected to have almost identical colors, we use this to estimate the CSPN 
reddening $E(B - V) = (B- V) - (B- V)_0$. Hence, the $B - V = -0.092$
CSPN color implies a reddening of $E(B - V) = 0.24 \pm 0.03$, in excellent 
agreement with the cluster weighted average from the available literature of $E(B - V) = 0.26 \pm 0.04$. The Gaia 
magnitudes are reported in \citet{2022arXiv220703179G}. An 
interstellar extinction profile was derived from the 3-D galactic model of 
\citet{Vergely2022} retrieved
through EXPLORE\footnote{https://explore-platform.eu}. It is an approximately linear 
rising trend until about 500pc, has a shallower slope then to 1000pc
with A$_o$ of $\sim$0.5 and flattens thereafter, rising only to A$_o$ of $\sim$0.57 by 
2000~pc. The main interstellar clouds are at 200~pc, 400~pc and 1~kpc, with another 
cloud at 1.6~kpc just behind M37. There is no extinction associated with the cluster
itself. This also supports interpretation of the nebula as a PN rather than ionized 
interstellar material.

\section{M37 cluster CMD, CSPN properties and the initial-to-final mass relation}
An M37 cluster CMD ($B$ versus $B-R$) diagram is shown in Fig.\,4, as 
generated from Gaia DR3 data and fit with an appropriate Padova 
theoretical isochrone for the adopted cluster parameters (age = 470~$\pm$~50 
Myrs, reddening $E(B-V)$ = 0.26$~\pm$~0.04, distance = 
1.49~$\pm$~0.13 kpc and metallicity $[Fe/H]$ = 0.03~$\pm$~0.28) as the yellow track; 
see \citet{2012MNRAS.427..127B} and \citet{2013MNRAS.434..488M}. The CSPN is 
shown by a red filled symbol and falls where WD cluster members are expected.  Stars 
with  $>$80\% probability \citep{2018A&A...618A..93C} of being a cluster member are 
plotted as green dots. The CMD includes all stars with pmRA = 0 to 4 and pmDec= $-8$ to
$-2$ mas/yr (most probable cluster members based on mean proper motions) within 
15~arcminutes from the cluster's apparent center.

%\begin{figure}[h!]
\begin{figure}
\begin{center}
\includegraphics[width=15cm]{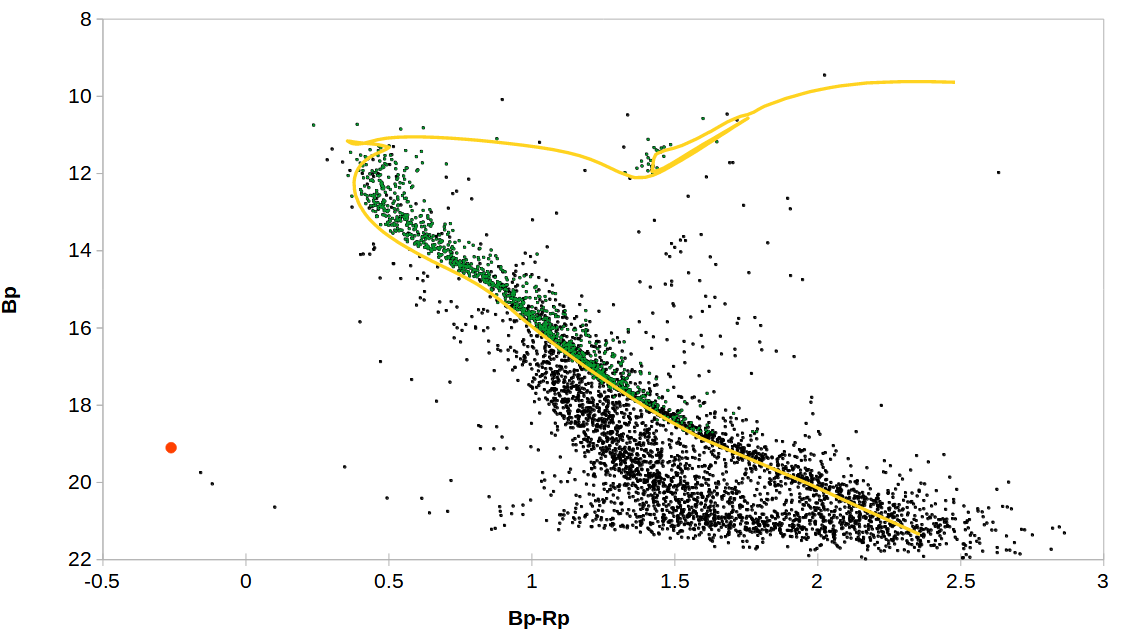}
\end{center}
\caption{Cluster Gaia DR3 CMD (B versus B-R) diagram fitted with a Padova 
theoretical isochrone (\citet{2012MNRAS.427..127B}, 
\citet{2013MNRAS.434..488M}
for adopted cluster parameters (age = 470~$\pm$~50 Myrs, reddening $E(B-V)$ = 0.26$~\pm$~0.04, distance = 
1.49~$\pm$~0.13 kpc and metallicity $[Fe/H]$ = 0.03~$\pm$~0.28)}. The CSPN is indicated by the red filled 
symbol. Stars with $>$80\% probability \citep{2018A&A...618A..93C} of being a cluster member,
where cross-correlated with Gaia DR3 and are plotted as green dots. The CMD includes all 
stars with pmRA =0 to 4 and pmDec = $-8$ to $-2$\,mas/yr (most probable cluster members based on mean proper motions) within 15~arcminutes from the cluster's apparent 
center.\label{fig:CDM}
\end{figure}

\subsection{CSPN derived parameters}
From the adopted cluster parameters and considering the time for the star to leave the Main 
Sequence and pass through the RGB/AGB phase, the adopted isochrone predicts a PN progenitor 
mass of 2.78$_{-0.1}^{+0.12}$ M$_\sun$. 

For the adopted CSPN $V$ magnitude, cluster reddening and distance, the CSPN 
absolute $V$ magnitude is $M_V = 7.56 \pm 0.19$. By plotting the CSPN $V$ 
magnitude and nebular kinematic age along with evolutionary tracks (see e.g. 
\cite{2009JPhCS.172a2033W}, their Figure 1, but using the 
\cite{2016A&A...588A..25M} improved tracks), we estimate a CSPN final mass of 
0.63$_{-0.04}^{+0.03}$ M$_\odot$ which is well within the expected range for a
WD descended from a progenitor star of $\sim$2.8~M$_\odot$.

The CSPN absolute $V$ magnitude and a series of possible CSPN effective temperature values (50
-- 300 kK, in steps of 10 kK) were used to calculate a series of CSPN luminosities,
which were plotted along with the corresponding evolutionary track 
\citep{2016A&A...588A..25M}. From the intersection of our line with the evolutionary tracks 
we estimate a CSPN effective temperature of 100 $\pm$ 20~kK (assuming 20\% error). This leads
to a CSPN luminosity estimate of $logL/L_\sun$ = 1.49 $\pm$ 0.25. 

\subsection{The IFMR}
The known sample of cluster WDs for the latest 
IFMR estimates and semi-empirical ‘PARSEC’ fit \citep{2018ApJ...866...21C} is 
presented in Fig.5. Our new estimate for the initial and final mass for our M37 cluster PN
IPHAS J055226.2+323724 is plotted as a red circle. The other 
two colored points are for the other known Galactic OC PNe; PHR~1315-6555 
\citet{2011MNRAS.413.1835P} and \citet{2019MNRAS.484.3078F} plotted as a 
yellow circle and the very high mass progenitor of PN 
BMP~J1613-5406, \citet{2019NatAs...3..851F} plotted as a green circle. The errors on each point reflect errors in the adopted cluster parameters and the
spread of the estimated CSPN magnitudes. 

\begin{figure}[h!]
\begin{center}
\includegraphics[width=10cm]{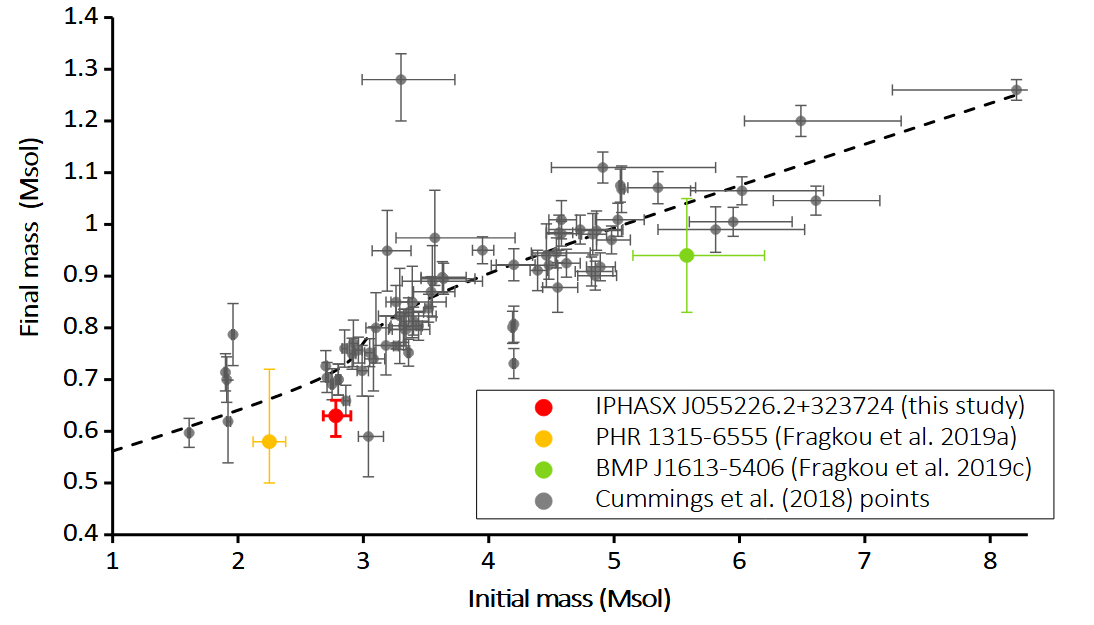}
\includegraphics[width=7.5cm, height=6cm]{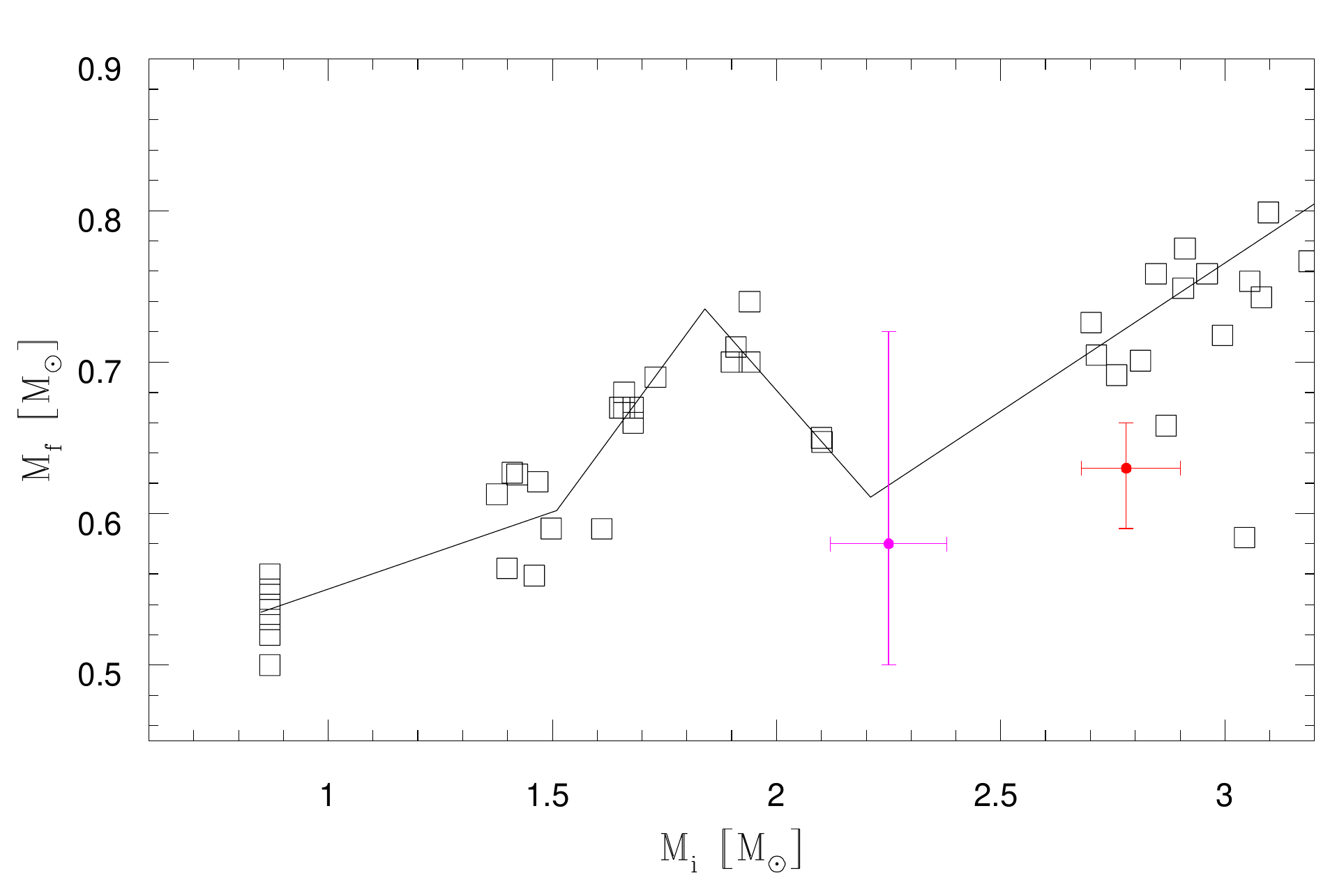}
\end{center}
\caption{Left panel: A plot from the known sample of cluster white dwarfs for the 
latest IFMR estimates and semi-empirical ‘PARSEC’ fit 
\citep{2018ApJ...866...21C} together with our estimated point for PN
IPHAS J055226.2+323724 plotted as a red circle. The other two points from 
known open-cluster PNe are plotted as a yellow circle (PHR~1315-6555 
\citep{2019MNRAS.484.3078F} and \citep{2011MNRAS.413.1835P}) and green circle 
(BMP~J1613-5406 - \citet{2019NatAs...3..851F}). The 
errors attached
to our point reflect the errors in the adopted cluster parameters and the 
spread of the estimated central-star magnitudes. Right panel: The initial final mass relation
of \citet{2020NatAs...4.1102M} with the two cluster PNe within this mass 
range}\label{fig:IFMR}
\end{figure}
%A straight line can be plotted through the 3 coloured points determined just from 
The OC PNe fall below the plotted dashed line trend from the cluster WDs, which barely 
overlap with the error bars. Recently \citet{2020NatAs...4.1102M} found an increase 
in final masses for initial masses in the range 1.5--2 M$_\odot$, based on white dwarfs
in the clusters R~137 and NGC~7789. Beyond this excursion the final masses fall 
back to the previous relation. The putative `kink' is proposed to be related to 
carbon star formation on the Asymptotic Giant Branch. The interpretation is 
complicated by  having only a single WD star in the range 2--2.5 M$_\odot$ from cluster 
NGC~752. The location of the two lower initial mass OC CSPN, the cluster white dwarfs 
and the proposed relation of \citet{2020NatAs...4.1102M} from their Fig.1 are 
shown in our version as Fig.\,5b. Their errors in M$_f$ are $\sim$0.5~$M_\sun$ but only 
$\sim$0.05~$M_\sun$ in M$_i$. Both cluster CSPN trace the key mass range just above 
the kink and in fact accentuate it, with one in the middle of the empty
mass range 2--2.7\,M$_\odot$. These cluster CSPN are 
consistent with the proposed reduction in this mass range from \citet{2020NatAs...4.1102M}.

%We hope our work in preparation on two further OC-PN candidates could add directly to this potentially interesting finding.

\begin{table}
	\centering
	\caption{Summary table of all determined PN, CSPN and cluster parameters from this work and from the literature.}
	\label{tab:coord}
	\begin{tabular}{lcc}
	\hline
         Parameter & PN/CSPN & Cluster M37 \\ %\hline \\
         RA (J2000) & 05:52:26.191 & 05:52:18.00 \\
         DEC (J2000) & 32:37:24.89 & 32:33:12.00 \\
         Apparent diameter (arcmin) & 7.42 & 31.15 \\
         Radial Velocity & 10.61\footnote{median of all 
         nebular MEGARA exposures}, $\sigma$ = 4.93 km/s & 8.32 $\pm$ 0.56
         km/s \\
         Distance & same as cluster\footnote{We assume the identified blue 
         star is the CSPN} & 1.49 $\pm$ 0.11 kpc\\
         Expansion velocity & 20, \footnote{calculated from the split of the  [\ion{N}{2}]
         6584 line} $\sigma$ = 6.2 km/s &\\
         Reddening $E(B-V)$ & 0.24 $\pm$ 0.03 \footnote{as 
         estimated from the CSPN colors} & 0.26 $\pm$ 0.04 \\
         Electron density $N_e$ & 5 cm$^{-3}$ \footnote{as 
         implied by the median of the  [\ion{S}{2}] 6716/6731 line 
         ratios of the four nebular MEGARA exposures acquired
         on March 4th 2022} &  \\
         Physical radius & 1.60 pc & 6.73 pc \\
         Morphology & Bipolar & open cluster \\
         Chemistry & likely Type I & $[Fe/H]$= 0.03 $\pm$ 0.28 \\
         Age & $78\times 10^3\pm25\times 10^3$~yrs & 470 $\pm$ 50 Myr  \\
         Estimated nebular mass & 0.32 M$_\sun$ & \\
         Estimated CSPN T$_{eff}$ & 100 $\pm$ 20 kK & \\
         Estimated CSPN Luminosity 
         $logL/L_\sun$& 1.49 $\pm$ 0.25 & \\
         
         Proper motion pmRA (mas/yr)  \footnote{PN proper motions refer to Gaia DR3 data for the 
         identified CS, while cluster proper motions refer to Gaia 
         DR3 median of cluster members with 
         membership probability $>$ 0.8 from 
         Cantat-Gaudin et al. (2018). }  & 2.00 $\pm$
         0.39  & 1.88  with 
         $\sigma$ = 0.18  \\
         Proper motion pmDec (mas/yr) & -5.37 
         $\pm$ 0.22  & 
         -5.62 with 
         $\sigma$ = 0.16  \\
         
         Central star $V$ magnitude & 19.16 $\pm$ 0.02 & \\
         Central star absolute magnitude $M_V$ & 7.56 $\pm$ 0.19 \footnote{as derived for the adopted cluster distance and reddening}  & \\
         Central star initial mass &  2.78$_{-0.1}^{+0.12}$ M$_\sun$ & \\
         Central star  final mass &  0.63$_{-0.04}^{+0.03}$ M$_\sun$ &\\ 
         Central star of PHR 1315-6555 initial mass\footnote{initial and final mass values for the other 2 OC-PN are taken from \cite{2019MNRAS.484.3078F}} &  2.25$\pm 0.13$ M$_\sun$ & \\
         Central star of PHR 1315-6555 final mass &  0.58$_{-0.08}^{+0.14}$ M$_\sun$ &\\
         Central star of BMP J1613-5406 initial mass &  5.58$_{-0.43}^{+0.62}$ M$_\sun$ & \\
         Central star of BMP J1613-5406 final mass\ &  0.94$\pm 0.11$ M$_\sun$ &\\ %\\\hline
	\end{tabular}
\end{table}

%\begin{figure}[h!]
%\begin{center}
%\includegraphics[width=8cm]{Mar.pdf}
%\caption{The initial final mass relation of %\citet{2020NatAs...4.1102M} with the two cluster PNe %within this mass range
%}\label{fig:pm}
%\end{center}
%\end{figure}

%\ref{fig:pm}. 

\section{Discussion and Conclusions}
Confirming an association of a PN and a star cluster requires close
agreement of multiple parameters for the PN, CSPN and cluster. These include PN positional proximity within the cluster tidal radius, reddening 
and distance agreement, a plausible CSPN, consistent PN and CSPN parameters, 
and crucially, radial velocity concurrence and proper motion agreement. OC 
velocity dispersions are typically $\sim$1~km/s so agreement here is a 
particularly tight constraint. The CSPN of IPHASX~J055226.2+32372 passes all 
these tests and crucially the radial velocity is in excellent agreement to 
within the errors. There is therefore high confidence that the CSPN and its PN reside in 
the M37 cluster, one of only three known physical associations of a PN with an OC in our Galaxy.

All results for the PN IPHASX~J055226.2+32372, CSPN and M37 cluster 
estimated here and those summarised from the available 
literature are provided  conveniently in Table.~1. All key conditions 
necessary to associate the PN and CSPN with the cluster are shown to be 
well satisfied. The kinematic age of 78~Kyrs appears to be the 
largest ever determined, assuming the measured expansion velocity has 
remained invariant with time. 

The three Galactic OC PNe come from relatively high (and in one case, very high) progenitor masses. They have notable commonalities. All 
are bipolar and appear to have Type-I chemistry
and high  [\ion{N}{2}]/H$\alpha$ ratios.  In the two cases of higher progenitor 
mass, the PNe are highly evolved and physically very large. As higher mass stars evolve 
quickly through the PN phase to enter the white dwarf cooling track this naturally leads to 
faint central stars and fainter nebulae. These open clusters have now revealed  
such nebulae which are very rare among the known field PNe. 

The large PN age found raises the issue of the maximum observable life time of PNe in general
\citet{Wareing2007}. Any PN shell is limited by interaction with the
interstellar medium (ISM). By the time the shell has swept up more mass than its own
ejecta, momentum conservation will cause it to adopt the ISM velocity while the 
CSPN moves with its own peculiar velocity and can leave the PN behind. The nebula eventually
looses its structure and dissipates into the ambient ISM. The 3 Galactic OC-PN uncovered so far all 
appear to avoid this fate. Stars in OCs have very little velocity dispersion with a systemic 
velocity likely very close to that of the ISM itself. This limits PN disruption so it is 
conceivable that OC PNe can be seen for longer than those 
in the field. In the case of the M37 PN, the brightening of the shell towards 
the south may herald the interaction with the ISM that will eventually disrupt it. 

%The estimated PN mass of $\sim 0.3$M$_\odot$ is only a fraction of the mass lost by the star, of around 2.5\,M$_\odot$. The remainder must have been lost before the mass was ejected that forms the PN. At a typical final mass loss rate of $\dot M \approx 2 \times 10^{-5}\,\rm M_\odot\,y^{-1}$ \citep{Decin2019}, the PN represents $1.5 \times 10^4$\yr. The earlier mass loss is expected to form a distant halo.

%As noted a straight line can  be fit through the 3 OC-PN points in the IFMR in Fig.5. which also falls just below the main trend found for cluster WDs. 

%We have several other OC-PN examples still in preparation that we hope  will clarify this possible linear relation further as the OC-PN sample have tighter constraints  on the initial mass because they have just recently descended from the  cluster MS turn-off. This is unlike cluster WDs whose time on the  cooling track is very difficult to estimate. 

OC PNe give important IFMR constraints. All three known Galactic examples 
fall just below the established relation \citep{2018ApJ...866...21C}. One traces the 
WD high mass end and one falls in the gap around 2--2.5\,M$_\odot$. The PNe show that 
the proposed steepening of the increase of WD masses seen at $M_i \sim 1.6\,\rm 
M_\odot$ to 1.9  \citep{2020NatAs...4.1102M} does not continue towards higher masses but 
drops back to the previous level, consistent with that paper's model.

\begin{acknowledgments}
\section*{Acknowledgements}
VF is in receipt of a UNAM postdoctoral fellowship. This work was 
supported by UNAM-PAPIIT grant IN106720. QAP thanks the Hong Kong Research 
Grants Council for GRF research support under grants 17326116 and 17300417. LS acknowledges UNAM-PAPIIT grant IN110122. 
This publication is based on data obtained with the MEGARA instrument at the 
GTC, installed in the Spanish Observatorio del Roque de los Muchachos, in the 
island of La Palma under program GTC1/21BMEX. MEGARA has been built by a 
Consortium led by the Universidad Complutense de Madrid (Spain) and the 
Instituto de Astrofısica, Optica y Electronica (Mexico), Instituto de 
Astrofısica de Andalucıa (CSIC, Spain), and the Universidad Politecnica de 
Madrid (Spain). MEGARA is funded by the Consortium institutions, GRANTECAN 
S.A. and European Regional Development Funds (ERDF), through Programa 
Operativo Canarias FEDER 2014-2020.  This work made use of the University of 
Hong Kong/Australian Astronomical Observatory/Strasbourg Observatory H-alpha 
Planetary Nebula (HASH PN) database, hosted by the Laboratory for Space 
Research at the University of Hong Kong; the extinction map query page, hosted
by the Centre for Astrophysics and Planetary Science at the University of 
Kent; and of data products from the IPHAS survey. This research has made use 
of the SIMBAD data base and the VizieR catalogue access tool, CDS, Strasbourg,
France (doi: 10.26093/cds/vizier). The original description of the VizieR 
service was published in Ochsenbein, Bauer \& Marcout (2000). This work has 
made use of data from the European Space Agency (ESA) mission Gaia 
(https://www.cosmos.esa.int/gaia), processed by the Gaia Data Processing and 
Analysis Consortium (DPAC, 
https://www.cosmos.esa.int/web/gaia/dpac/consortium). This research has used data, tools or materials developed as part of the EXPLORE project that has received funding from the European Union’s Horizon 2020 research and innovation programme under grant agreement No 101004214. We gratefully 
acknowledge the earlier work of our colleague Dr. David Frew who first brought
this important candidate to our attention in 2017. We thank our colleague Dr. Ivan Bojicic for his help with the creation of the cluster mosaic. We thank the anonymous referee for a careful and efficient review that improved the paper.
\end{acknowledgments}

%% To help institutions obtain information on the effectiveness of their 
%% telescopes the AAS Journals has created a group of keywords for telescope 
%% facilities.
%
%% Following the acknowledgments section, use the following syntax and the
%% \facility{} or \facilities{} macros to list the keywords of facilities used 
%% in the research for the paper.  Each keyword is check against the master 
%% list during copy editing.  Individual instruments can be provided in 
%% parentheses, after the keyword, but they are not verified.

\vspace{5mm}
\facilities{ GTC:10.4m,MEGARA; INT 2.5m \& IPHAS}

%% Similar to \facility{}, there is the optional \software command to allow 
%% authors a place to specify which programs were used during the creation of 
%% the manuscript. Authors should list each code and include either a
%% citation or url to the code inside ()s when available.

\software{IRAF \citep{1986SPIE..627..733T, 1993ASPC...52..173T}, MEGARA pipeline \citep{2022zndo....593647P} 
          }

%% Appendix material should be preceded with a single \appendix command.
%% There should be a \section command for each appendix. Mark appendix
%% subsections with the same markup you use in the main body of the paper.

%% Each Appendix (indicated with \section) will be lettered A, B, C, etc.
%% The equation counter will reset when it encounters the \appendix
%% command and will number appendix equations (A1), (A2), etc. The
%% Figure and Table counter will not reset.

%% For this sample we use BibTeX plus aasjournals.bst to generate the
%% the bibliography. The sample631.bib file was populated from ADS. To
%% get the citations to show in the compiled file do the following:
%%
%% pdflatex sample631.tex
%% bibtext sample631
%% pdflatex sample631.tex
%% pdflatex sample631.tex

\bibliography{bibliography}{}
\bibliographystyle{aasjournal}

%% This command is needed to show the entire author+affiliation list when
%% the collaboration and author truncation commands are used.  It has to
%% go at the end of the manuscript.https://www.overleaf.com/project/62cc1e417678fc66c1391e7d
%\allauthors

%% Include this line if you are using the \added, \replaced, \deleted
%% commands to see a summary list of all changes at the end of the article.
%\listofchanges

\end{document}